\documentclass[preprint,showpacs,aps]{revtex4}
\usepackage{graphicx}

\begin{document}

\title{Semiclassical wave packet tunneling in real-time}  

\author{Joachim Ankerhold}
\email{ankerhold@physik.uni-freiburg.de}
\author{Markus Saltzer}
\affiliation{
Physikalisches Institut, Albert-Ludwigs-Universit{\"a}t
Freiburg, Hermann-Herder-Stra{\ss}e 3, D-79104 Freiburg, Germany}

\date{\today}
           

\begin{abstract}
Quantum mechanical real-time tunneling through general scattering potentials
is studied in the semiclassical limit. It is shown
that the exact path integral of the  real-time propagator is
dominated in the long time sector  by quasi-stationary
fluctuations associated with caustics. This leads to an extended semiclassical 
 propagation scheme for wave packet dynamics 
 which accurately describes 
 deep tunneling through static and, for the first time,
driven barrier potentials.  

\end{abstract}

\maketitle

\section{Introduction}
Tunneling through a potential barrier is one of the most fascinating
aspects of quantum mechanics. In recent years a particular challenge
has been to understand tunneling in complex systems using
semiclassical methods. 
 However, 
any simple description is seemingly
hampered by the fact that a quantum mechanical object
running towards a barrier 
with a typical energy smaller than the barrier height may
penetrate it even though all classical trajectories with such
energies are reflected completely. 
Thus, for static observables like e.g.\ tunnel splittings one works in the
{\em energy domain} and calculates the energy dependent
Green's function semiclassically by 
switching from a 
real-time  orbit outside the barrier 
 to an orbit in imaginary time, i.e.\ with imaginary momentum,  under the
barrier. This technique can be traced back to the ``old'' WKB
approximation \cite{landau} and meanwhile has been successfully extended to
extract e.g.\ tunnel splittings also for systems with classically
chaotic dynamics \cite{creag}.
The crucial question is then: Can semiclassical tunneling also be described in
the {\em real-time domain}? This issue has turned into a fundamental challenge
for our understanding of semiclassics in general and systems with
explicit time dependence as e.g.\ in the context of driven tunneling
and control of tunneling \cite{grifoni} or tunneling in the presence of chaos
\cite{chaos} in particular.

The probability amplitude for a particle initially at $q_i$ to be at
$q_f$ after time $t$ is given by Feynman's path integral representation
of the propagator as \cite{schulman}
\begin{equation}
G(q_f,q_i,t)\equiv \langle q_f|{\rm e}^{-i H t/\hbar}|q_i\rangle
= \int {\cal D}[q]\, {\rm e}^{i S[q]/\hbar}.\label{eq1}
\end{equation}
The integral sums over all paths running from $q_i$ to $q_f$ in time $t$
where each contribution is weighted according to its action
$S[q]=\int_{q_i}^{q_f} dq \sqrt{2m[E-V(q)]}- E t$ with potential $V(q)$ 
 and energy $E=E(q_f,q_i,t)$. Due to the oscillating integrand in
(\ref{eq1})  tunneling appears
to be the result of a complex interference pattern. In the semiclassical
limit the path sum is dominated
by the contributions of the  stationary paths $\delta S[q_{\rm cl}]=0$
obeying Newton's equation of motion 
and  small fluctuations around them.
In the last decade efficient semiclassical propagation schemes
based on Gaussian wave packets have been developed, certainly
the most powerful known as the Hermann-Kluk propagator (HK) \cite{kluk}. 
However, the inclusion of deep tunneling (may be even
in presence of external driving) has not
been satisfactory yet. In fact, it was found that classically allowed real-time
trajectories running {\em over} the barrier are not
sufficient to capture strong tunneling \cite{heller2,heller1}. Various
extensions have 
thus been attempted. By propagating a large number of initial wave
packets tunnel splittings in a
double well potential could be 
extracted, however,  in single wave packet motion
tunneling effects were absent \cite{mandel}. As for  barrier penetration no
low energy 
stationary orbit exists along 
the real axis, it is tempting to think that one may find one in 
the complex plane \cite{heller1}. An individual path
circumventing--in 
real-time--the barrier region in a complex coordinate plane
is assumed, but its existence is still
elusive. In \cite{grossmann} orbits run  along the
real axis, but the semiclassical propagation over the time interval
$t$ is time sliced into steps over intermediate intervals.  This
spawning of orbits turns out  to be numerically extremely expensive already for
two slices and improves tunneling amplitudes only for
energies not too far below the barrier top. 

In the sequel we re-examine the semiclassical barrier penetration
through scattering potentials starting from the exact expression
(\ref{eq1}). Our idea is this: While in the energy domain
tunneling is described 
within a complex time plane, here, we 
study classical mechanics for complex energies. 
Our analysis reveals how tunneling is encoded in
the quantum propagator (\ref{eq1}) in terms of real-time orbits. In
particular, it 
turns out that an individual complex ``tunneling path'' does not
exist. 
Based on these results we extend the conventional HK to semiclassical
wave packet 
dynamics in the deep tunneling 
regime not only through static
 potentials, but, for the first time, report also on accurate results
for driven tunneling. The approach is shown to be very efficient for
one-dimensional systems and may thus serve as a promising starting point for
higher dimensional studies.

\section{Complex mechanics and semiclassical approximation}
We consider the motion of a particle of mass
$m=1$ in a general
one-dimensional, symmetric barrier potential $V(q)$ where the barrier
top is located at $q=0$ with $V(0)=V_0$. $V(q)$ is assumed to be 
a smooth and analytic function of $q$ that can be approximated around
$q=0$ by an inverted harmonic oscillator and for large $|q|$ falls off
as $V(q)\to V_0/[q/l]^{2k}, k\geq 2$, integer, with a typical
barrier length scale $l$.  A sufficiently high barrier is taken for
granted.  Typical examples include $V_k(q)=V_0/[1+(q/l)^2]^k$, but 
our results also apply to the Eckart
barrier $V_0/\cosh(q/l)^2$ and the Gaussian barrier $V_0
\exp(-q^2/l^2)$. 

Now, think of a wave packet $\psi(q_i,0)$ localized to the far right
($q_i>0$) which is propagated towards the barrier according to 
$\psi(q_f,t)=\int dq_i G(q_f,q_i,t) \psi(q_i,0)$. We are interested in
that portion of the packet that after time $t$ arrives on the far
left ($q_f=-q_i<0$). All
real orbits connecting the two asymptotic regions run over the barrier
($E>V_0$) and as, for fixed 
end-points, $t$ 
becomes large they spend most of their time in the parabolic
range around $q=0$. There, the marginal stability of trajectories causes
 the semiclassical $G(-q_i,q_i,t)$  to die out
exponentially in contrast to exact results \cite{heller2}. 
Classical orbits with  $E< V_0$ coming from the
far right or left 
reach the right or left flank of the barrier at turning points (TPs)
$q_0$ and $-q_0$, respectively; the long time properties of the path integral
(\ref{eq1}) are therefore governed by the dynamics in the
``forbidden'' range between 
the TPs. 
Mathematically, for $E<V_0$ no {\em real} stationary phase point to
(\ref{eq1}) obeying the proper boundary conditions
exists in function space. The usual procedure is then an analytic
continuation meaning here to extend classical mechanics to the
complex coordinate plane.

Newton's equation of motion, $\ddot{q}+V'(q)=0$, where $\dot{q}=dq/dt$
and $V'=d
V/dq$, translates for complex $q=x+i y$ into
\begin{equation}
\ddot{x}+r_x=0\ ,\ \ \ \ddot{y}+j_x=0.\label{motion}
\end{equation}
Here, $V(q)=r(x,y)+i j(x,y)$ and the subscript $x\,$ [$y$]
denotes the partial derivative with respect to $x\, $ [$y$]. We
further exploited that for analytic functions $V(q)$ Cauchy's relations
$r_x=j_y$ and $r_y=-j_x$ apply. From (\ref{motion}) one simply finds that
 the total energy $E=\epsilon_{\rm re}+i \epsilon_{\rm im}$
and 
its real and imaginary parts 
\begin{equation}
\epsilon_{\rm
re}=(\dot{x}^2-\dot{y}^2)/2+r(x,y)\ \ \ \mbox{ and}\ \ \ \  \epsilon_{\rm
im}=\dot{x}\dot{y}+j(x,y)\, ,\label{ener}
\end{equation}
respectively, are constants of motion. 
How does the corresponding classical mechanics look like?
For low energies the TPs $q_0, -q_0$ lie in the range
where $V(q)$ can be approximated by its asymptotic behavior. Hence, 
we consider paths starting from large $q_i=x_i>0$ along the real axis with
complex momentum $p_i\equiv\dot{q}_i=\dot{x}_i+i \dot{y}_i$. Typical
trajectories are 
depicted in fig.~\ref{fig1}. While basically three kinds of orbits can be
distinguished, the common behavior  is that  as the barrier 
vanishes asymptotically, for 
large distances from the top the classical motion tends to be
a free motion. If we  represent trajectories in the form
$q(t)= R(t) {\rm e}^{i \phi(t)}$, for very large $R$ they
 run close to straight lines with constant $\phi(q_i,p_i)$ depending merely
on the initial phase space variables. Let us now discuss the types of
orbits in detail.

First, we look at class (a) as it is the only one where orbits
cross the line $x=0$ ($\phi=\pi/2$) to reach the other side
of the barrier. For such 
orbits we need initially  $\dot{x}_i<0$ (otherwise they would run
away from the barrier anyway) and for the discussion
assume $\dot{y}_i>0$ implying $\epsilon_{\rm
im}=\dot{x}_i\dot{y}_i<0$, see Eq.~(\ref{ener}). We further 
note the asymptotic form of $V(q)=r(q)+i j(q)$ using polar coordinates:  
\begin{equation}
r(R,\phi)= \frac{\cos( 2 k \phi)}{(R/l)^{2k}} \ , \ \
j(R,\phi)=-\frac{\sin(2 k \phi)}{(R/l)^{2k}} \, .\label{poten}
\end{equation}
One sees immediately that a successful crossing of the dividing surface
 must happen with $\dot{y}>0$ (and $\dot{x}<0$ of
course). Namely, at $\phi=\pi/2$ the imaginary
part $j$ vanishes so that 
$\epsilon_{\rm im}=\dot{x}\dot{y}=\dot{x}_i\dot{y}_i<0$ meaning
$\dot{y}>0$. For the required
energy one derives $|\epsilon_{\rm im}|>r(x_i,0)=V(x_i)$.  Starting, 
however, from the imaginary axis with $\dot{x}<0$ and $\dot{y}>0$,
i.e.\ in the direction of decreasing $|V(q)|$, always generates an
orbit  reaching the asymptotic left side of the 
barrier far from the real axis ($\phi<\pi$). 
We conclude that
a simple ``tunneling path'' connecting the asymptotic segments of  the
real axis on either side of the barrier via a tour through the complex
plane does not exist.
This is in sharp contrast to tunneling for fixed energy. There, the  energy
dependent Greens function $K(q_f,q_i,E)$ exhibits stationary phase
points in imaginary time corresponding to classical paths running with
energy $E$ in
the inverted potential through the barrier range from $q_i$ to
$q_f$. Here, for real-time 
tunneling a stationary phase path to the quantum propagator
$G(q_f,q_i,t)$  even with complex energy
cannot be found. This important result may also reflect the
quite different roles ``energy'' and ``time'' play in quantum mechanics.

For our analysis the consequences are two-fold: on the one hand 
complex trajectories in class (a) do not play any role for a semiclassical
approximation to $G(q_f,q_i,t)$, and on the other hand the path integral in (\ref{eq1}) is in the low
energy sector completely determined by
fluctuations. To find its dominant contributions thus means to
detect the dominant fluctuations; these are points in function space which
lie close to orbits with $\delta S[q]=0$ and also obey
the proper boundary conditions. Accordingly, we consider the
remaining two classes of paths.

The second class  (b)
contains orbits with small but non-vanishing energies $0<|\epsilon_{\rm
im}|<r(x_i,0)$ which may exhibit TPs in the complex plane and always live
 on the same side of the barrier. Hence they are not relevant either.
In the third
class (c) trajectories have real total energy $E$, i.e.\ 
$\epsilon_{\rm im}=0$, but start with 
purely imaginary momenta 
$\dot{x}_i=0$ and, as assumed, small $\epsilon_{\rm
re}=-\dot{y}_i^2/2+r(x_i,0)$. 
These orbits display crucial features as we will
explain in the following. For that purpose we focus on the limit $\epsilon_{\rm
re}=0$ and follow paths with $x_i>0, \dot{y}_i>0$.
Writing asymptotically $q(t)= R(t) {\rm e}^{i \phi}$
one obtains 
\begin{equation}
\phi(x_i,\dot{y}_i)\equiv \phi_c^+=\pi/[2(k+1)]\, .\label{phas}
\end{equation}
Hence, after a
transient period of time 
 {\em all} those orbits run along the {\em same} line in the
complex plane independent of their starting points $x_i$. And since they 
carry the same energy, they are also focused in
phase space so that the line $\phi_c$ defines a caustic. Were the
trajectories  optical rays, $\phi=\phi_c^+$ would be a 
burning line. Due to symmetry the same holds true for the
complementary line $-\phi_c^+$ and the lines $\pi\pm\phi_c^+$ on the
other side of the barrier. 
Typically, a caustic is associated with
unstable orbits and fluctuations connecting them which renders a simple
Gaussian semiclassics insufficient \cite{schulman}. To verify 
this scenario here, we consider small deviations $\delta q= \delta x +
i\delta y$ around a 
certain orbit $\bar{q}(x_i;t)$. By linearizing the equations of 
motion (\ref{motion}) one gains
$(\delta\ddot{x},\delta\ddot{y})^T=${\boldmath $M$ \unboldmath }$
(\delta x,\delta y)^T$ where {\boldmath $M$ \unboldmath } is the
stability matrix 
evaluated along $\bar{q}(t)$. Along $\phi=\phi_c^+$ its diagonal elements are
$-\bar{r}_{xx}>0$ and the off-diagonal elements vanish 
$\bar{j}_{xx}=0$. Accordingly,  all trajectories merging along
the burning lines are unstable. Small
deviations in phase space can lead from an orbit $\bar{q}(x_i;t)$ to
another one $\bar{q}(x_i';t)$ and even allow for a turn from positive
 to negative momentum to  run along the
 $\bar{q}(x_i';t)$-orbit 
back towards the real axis. As asymptotically paths creep along
$\phi_c^+$, jumps from very small positive to 
negative momenta require only tiny fluctuations. 
The reversed orbit crosses the real axis at
$x_i'$ and  approaches the complementary burning line $-\phi_c^+$ 
 in the lower 
halfplane. There, a similar kind of deviation drives it to still
another $\bar{q}(x_i'';t)$ to reach again $\phi_c^+$ and so forth and
 back. By 
subsequently running through these cycles between the caustics at $\pm
 \phi_c^+$ a net-motion into the direction of the barrier
top may be generated. On the left side of the
barrier ($x_i<0$) the same kind of scenario exists and at the top 
$x=0$ the burning lines intersect (depending in detail on
$V(q)$ within $|q/l|\alt 1$, see fig~\ref{fig1}.) so that small deviations in
 the vicinity of the bottleneck $x=0$ may lead from the  set of
 right-barrier paths to that of left-barrier paths and vice versa.
 This allows for a motion starting in $x_i>0$ to
eventually reach the range on the opposite side of the barrier. So far
the above discussion is restricted to class (c)-orbits 
with $\epsilon_{\rm re}=0$. However, for 
 finite but small $\epsilon_{\rm re}$ orbits merge
 close to the burning lines, and we find basically the same situation. The
conclusion is that two real axis paths with
TPs at $q_0$ and $-q_0$, respectively, are linked by a sequence of
real-time complex plane orbits tied together by small
fluctuations near caustic lines. 
Since this under-barrier-motion is not a purely
 stationary one obeying (\ref{motion}), but can be seen as nearly
stationary as it follows classical orbits most of the time,
it describes quasi-stationary fluctuations (QSF).
The QSF allow to move from $q_i>0$ through the barrier
range towards $q_f<0$ and this way dominate in absence of true
stationary points, $\delta S[q]=0$ with $q(0)=q_i, q(t)=q_f$, the path
integration in $G(q_f,q_i,t)$ between the TPs.

\section{Extended semiclassical propagator}
The action of the QSF can simply be approximated. For a cycle from
$x_i$ with $\bar{q}(x_i;t)$ to $x_i'<x_i$ with $\bar{q}(x_i';t)$ in
the interval $\delta t$ we
find with Cauchy's formula $S(x_i',x_i,\delta t)\approx i
|W(x_i',x_i)|-E \delta t$ where
 the short action is $W(x',x)=
 \int_{x'}^{x} dq \sqrt{2m [E-V(q)]}$ and the portion from the
phase space deviation along $\phi_c^+$  is negligible. Accordingly, 
$S(-q_0,q_0,\Delta t)\approx i |W(-q_0,q_0)|-E \Delta t$
where $\Delta t$ is the time interval spent between the
real axis TPs $q_0,-q_0$. Hence, one arrives at the crucial result that the 
real-time motion of the QSF gives rise to an imaginary part
in the action which is identical to the known instanton or WKB
exponent. The full action for a low energy motion from $q_i>0$ to   
$-q_i$ now consists of two classical real axis segments from $q_i$
to $q_0$ and from $-q_0$ to $-q_i$, respectively,  and QSF
inbetween, i.e.\ $S(-q_i,q_i,t)\approx 2
W(q_0,q_i)+i|W(-q_0,q_0)|-E t$. In the semiclassical
$G(-q_i,q_i,t)$ the exponential of this action
is accompanied by the contribution of Gaussian fluctuations
around the real axis segments. 

The most powerful representation of the
semiclassical propagator is the so-called
Hermann-Kluk propagator (HK) \cite{kluk}. It has the advantage of
being determined 
by an initial value problem for the classical trajectories, namely,
\begin{equation}
G_{\rm HK}(q_f,q_i,t)=\int \frac{dq dp}{2 \pi \hbar}\,  h(q_f,q_i,t,p,q)\,
R(p,q,t)\,  
{\rm e}^{i S(p,q,t)/\hbar}  
\label{hk}
\end{equation}
with the fluctuation prefactor $R(p,q,t)$ and an overlap
factor $h(q_f,q_i,t,p,q)=\langle
q_f|\gamma(p,q,t)\rangle  \langle \gamma(p,q)|q_i\rangle$ where
\begin{equation}
\langle x|\gamma(p,q,t)\rangle=\left(\frac{\gamma}{\pi}\right)^{1/4}\,
\exp\left\{-\frac{\gamma}{2}\, [x-q(t)]^2+\frac{i}{\hbar} p(t)
[x-q(t)]\right\}\label{gauss}
\end{equation}
is a Gaussian wave packet centered around the
phase space point $\{p(t),q(t)\}$. In (\ref{hk}) one runs in the time interval
$t$ real trajectories 
from $\{p,q\}$ to $\{p(t),q(t)\}$  where the
contribution of each orbit is weighted according to its action and
fluctuation prefactor and the Gaussian  overlap of its end-points with
those of the propagator. 

The usual HK (\ref{hk}) is exact for a pure parabolic barrier and therefore
captures tunneling only for harmonic fluctuations around the barrier top but
 fails for long times \cite{heller1}. To overcome
this drawback we apply the results developed above and formally split the
propagator in phase space: 
$G(q_f,q_i,t)=G_>(q_f,q_i,t)+G_<(q_f,q_i,t)$ where $G_>$ [$G_<$]
contains orbits with $E\geq V_0$ [$E<V_0-\delta_{\rm pb}$] and
 $G_>$ coincides 
 with $G_{\rm HK}$ (\ref{hk}). Accordingly, $G_>(t)$ describes the time
evolution up to moderate times (comprising the  parabolic
range  $V_0-\delta_{\rm pb}$ below the top) and $G_<(t)$ the
 long time behavior. 
Now, while in a strict sense the complex dynamics discussed
above is only valid for very low $E$, we assume its
applicability also for somewhat larger $E$ and find 
with $q_i$ and $q_f$ on opposite sides of the barrier
\begin{equation}
G_<(q_f,q_i,t)=\hspace*{-0.5cm}\int\limits_{E<V_0-\delta_{\rm pb}}
\hspace*{-0.5cm} \frac{dq dp}{2\pi \hbar}\, h(q_f,q_i,t,p,q)\,  
R^<(p,q,t)\, {\rm e}^{i S^<(p,q,t)/\hbar}\, T(q_0).
\label{hkt} 
\end{equation}
Here, an
orbit runs from $\{p,q\}$  along the real axis to its TP $\{0,q_0\}$, jumps to
$\{0,-q_0\}$ to reach $\{p(t),q(t)\}$ leading to a fluctuation
prefactor $R_<$ and 
action $S_<$. The position space jump costs
$T(q_0)=\exp[-|W(-q_0,q_0)|/\hbar]$ and eventually $G_<$ results from
phase space averaging. In (\ref{hkt}) we require $|q_i|/l,
|q_f|/l\gg 1$ so that most of the dynamics
is spent outside the barrier. 
Obviously, $G_<$ follows {\em not} just from switching in the
integrand in $G_>$ to imaginary times in regions where $E<V(q)$.

\section{Applications}
The extended HK (eHK) $G_{\rm eHK}=G_>+G_<$ is now employed to
scattering in an Eckart 
barrier $V_b(q)=V_0/\cosh^2(q/l)$ 
that has been of wide  use, e.g.\ as a
model for the H+H$_2$ exchange reaction. Since asymptotically
$V(q)$ drops faster than any power of
$q$ we have $\phi_c^+\to 0$ and
burning lines stretch parallel to the real axis.
 In fig.~\ref{fig2}a the
correlation function 
$c_{fi}(t)=\langle \psi_f|\exp(-i H t/\hbar)|\psi_i\rangle$ between two
Gaussian wave packets is depicted. Initially, $\psi_i$ [$\psi_f$] is
centered to the far right [far left] with
$V_0\gg p_i^2/2$ so that we are indeed in a deep tunneling regime.
 One clearly
sees the exponential drop of $G_>$ and the 
startling accuracy of the eHK over the whole time range. 
The most sensitive observable for a
 real-time treatment is the transmission probability $P(E)$ calculated
by numerically Fourier 
transform $c_{fi}(t)$. 
Remarkably, we get accurate data also for very low energies
(fig.~\ref{fig3}) apart from small oscillations typical for real-time
calculations \cite{tannor}. In the moderate energy range
$E/V_0>0.5 $
the ``real-time'' $P(E)$  even improves
 the uniform WKB result. Here convergence for
$G_<$ is achieved for roughly the same number of trajectories as in
$G_>$ (typical number of trajectories for the set of parameters is
$5\cdot 10^4$) so that in contrast to previous approaches \cite{grossmann} an
extension of the 
eHK to two or three dimensional systems, which are of particular
interest to study chaotic tunneling, seems feasible.

As an example where a dynamical approach is clearly needed
 we turn to an Eckart barrier driven by a periodic
signal $V=V_b+q A \sin(\Omega t)$ and focus on the range
of non-resonant driving and weak to moderate
driving amplitudes. In this case already
the exact numerics is
non-trivial since it is the long time tunneling behavior which is 
 most sensitively  affected by the driving and leads to a 
strong spreading of the wave packet.
 Typical results for the correlation function $c_{fi}(t)$ are shown
in fig.~\ref{fig2}b. Compared to the static case one sees
 phase shifted oscillations and a revival type of phenomenon.
Semiclassically, both effects originate from an intimate
interference of (fast) above-barrier-paths ($E>V_0$), which cross the
barrier and then are back-scattered,  and (slow) driven tunneling orbits
($E<V_0$). Even in this 
time-dependent case the accuracy of the eHK is quite astonishing. 

\section{Conclusion}
To conclude, our findings reveal for the first
time how tunneling  is encoded in
 the quantum propagator in terms of classical real-time orbits.
In contrast to tunneling in the energy domain, real
time barrier penetration cannot be described by  individual
 tunneling orbits. Instead, it must be seen
as a diffusion along a certain set of classical paths in the complex
plane. 
This
allows for a practical approach for semiclassical wave packet 
dynamics even in the deep tunneling regime of 
static and non-resonantly driven scattering processes.
Explicit examples have been given for one dimensional cases, but the
efficiency of the method suggests that at least two or three
dimensional cases may be feasible. The main problem then will be
that from a certain TP $q_0$ a bunch of TPs on the other side of the barrier
can be reached. This 
proliferation of orbits, however,  seems tractable due to the exponential
suppression of under barrier motion starting at $q_0$ and traveling
over large distances. Work in this direction is in progress.
Possible applications of our method and extensions of it
in physics and chemistry may be e.g.\
 mesoscopic systems 
in microwave fields or unimolecular reactive scattering.

Financial support by the DFG through SFB276 is gratefully
acknowledged.

\pagebreak

\begin{figure}
\caption{\label{fig1}Orbits in the complex
plane for $V_2(q)=1/(1+q^2)^2$ [thin; class (a) and (b) dotted,
class (c) solid for various $x_i$ and 
$\dot{y}_i>0$, $\dot{y}_i<0$]. Burning lines (thick) are shown for 
$V_2(q)$ (dashed) and its asymptote $1/q^4$ (solid); dots are TPs.\hfill}
\end{figure}

\begin{figure}
\caption{\label{fig2}Real part of $c_{\rm
fi}$ vs.~time for the static (a) and driven (b) scattering in
an Eckart barrier. Parameters are $\gamma l^2=6$, $V_0/(p_i^2/2)=8$, 
and (a) $q_i/l=-q_f/l=40$, (b) $q_i/l=-q_f/l=15$ with $q_i
A/V_0=-0.75$, $\Omega/\sqrt{V_0/2l^2}=0.02$.\hfill}
\end{figure}

\begin{figure}
\caption{\label{fig3} Transmission probability vs.~$E/V_0$. Exact
(solid), usual HK (dotted), eHK (dashed), and uniform WKB
(dotted-dashed) are shown.}
\end{figure}

\begin{figure}
\center
\includegraphics[height=27cm,draft=false]{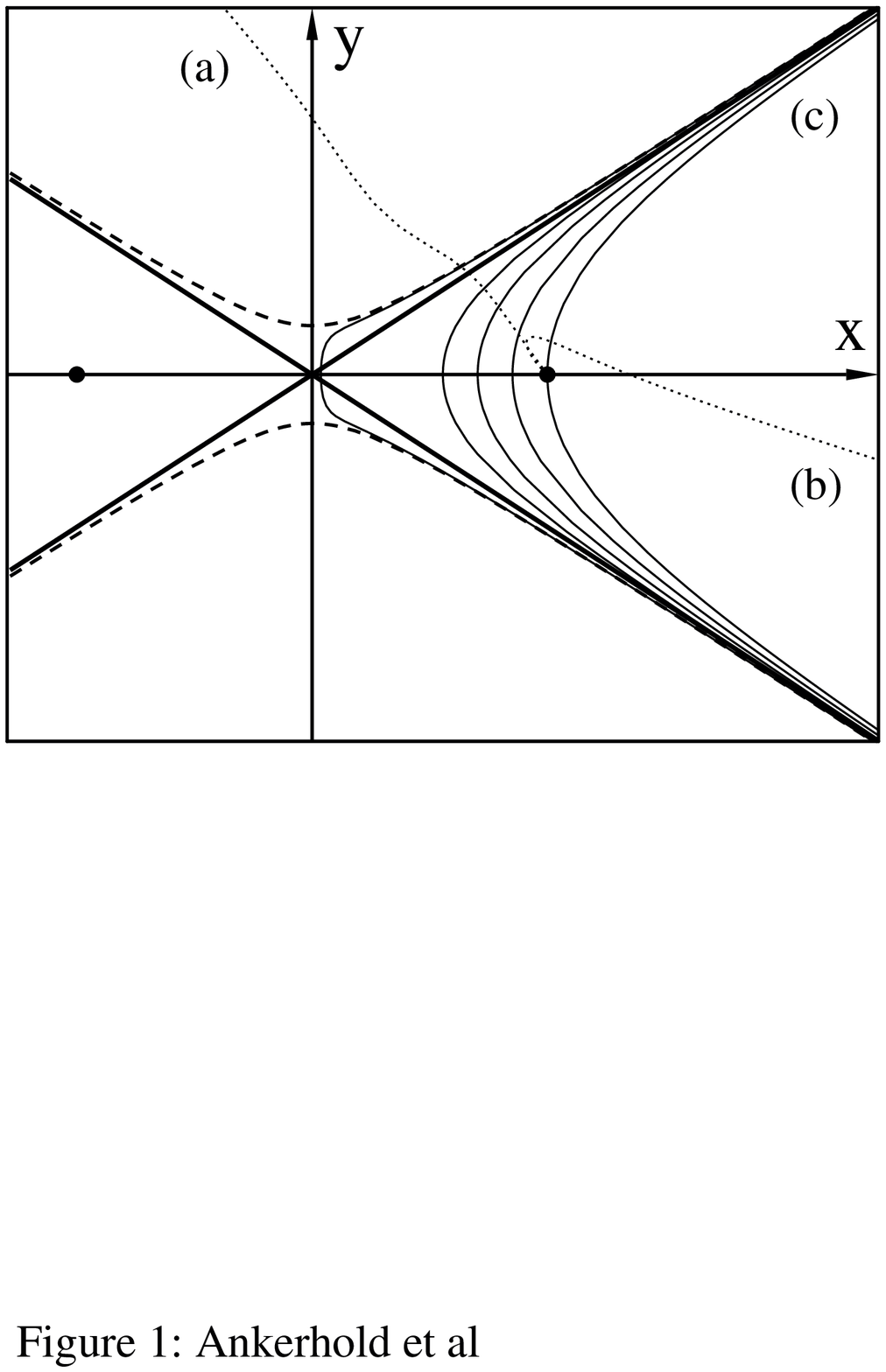}
\end{figure}

\begin{figure}
\center
\includegraphics[height=27cm,draft=false]{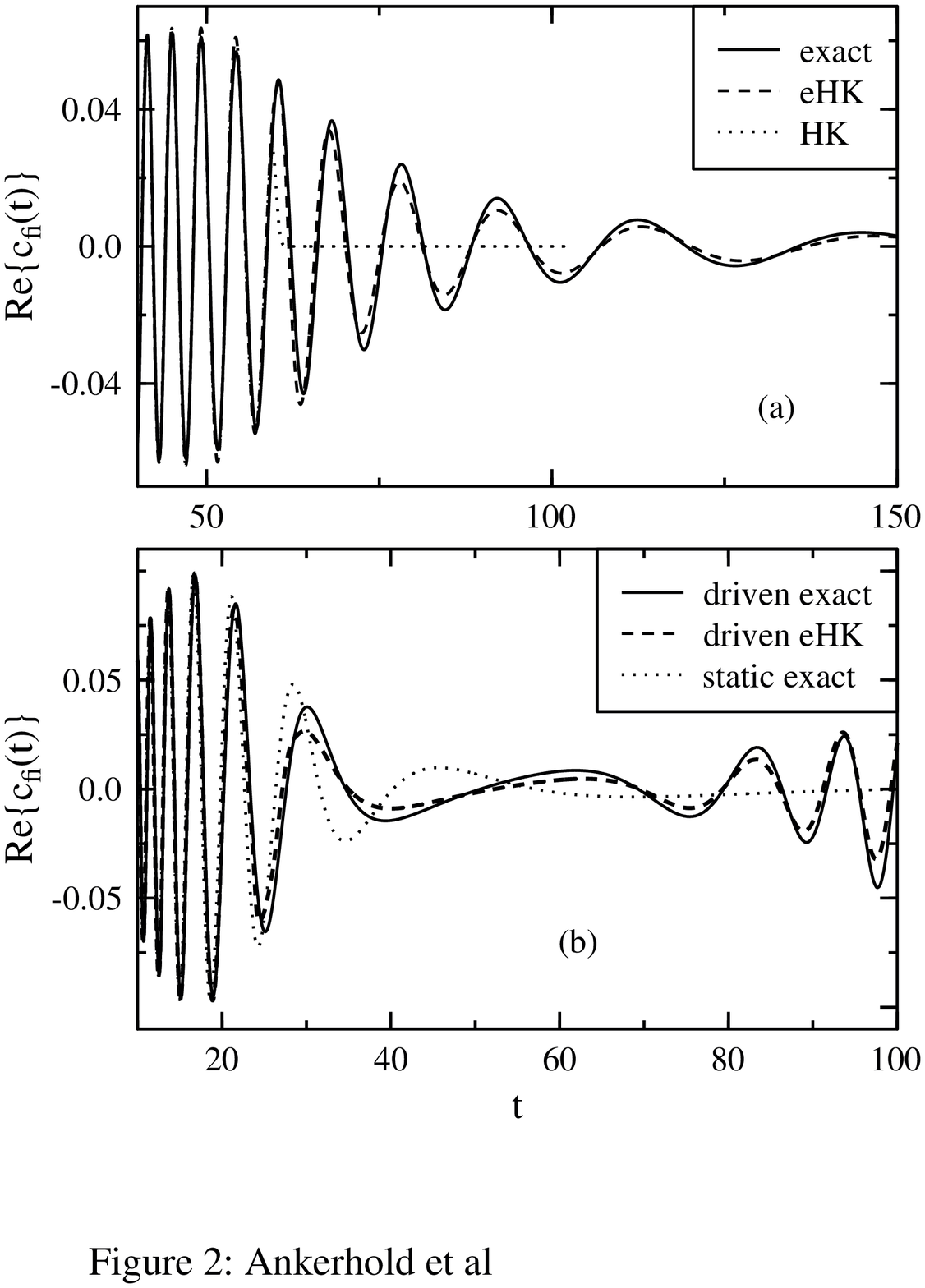}
\end{figure}

\begin{figure}
\center
\includegraphics[height=27cm,draft=false]{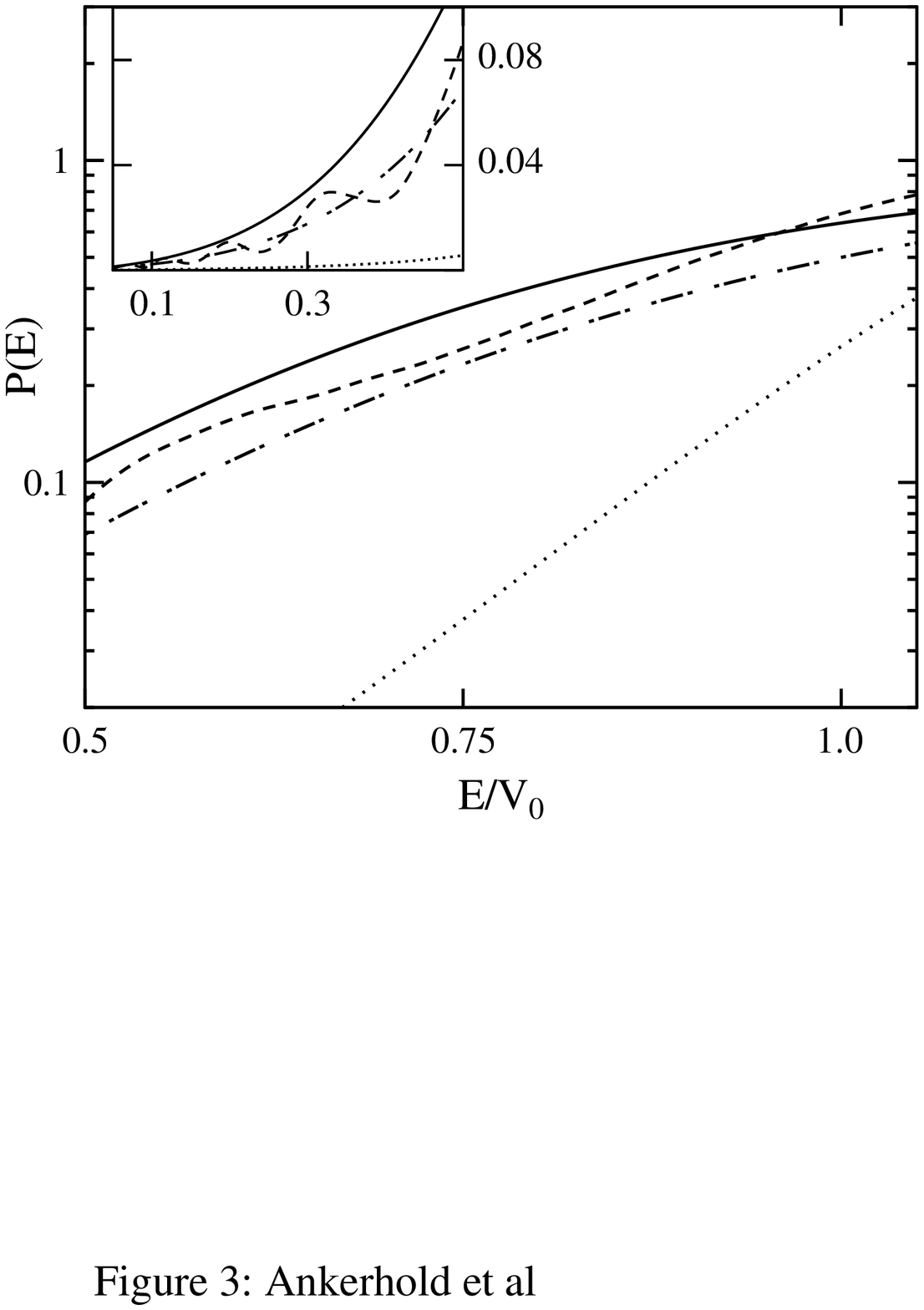}
\end{figure}

\end{document}